**Cuprate superconductors: Dynamic stabilization?**

*Enhancing the temperature at which superconductivity is observed is a long-standing objective for materials scientists. Recent tantalizing experiments suggest a possible route for achieving this.*

**N. Peter Armitage[1]**

[1] *N. Peter Armitage is at the Institute for Quantum Matter, Department of Physics and Astronomy, The Johns Hopkins University, Baltimore, Maryland 21218, USA*

For 25 years, researchers have tried a myriad of combinations to enhance the superconducting temperature of cuprate superconductors above the record of 134 K set for $HgBa_2Ca_2Cu_3O_8$ [1]. The hope has always been that just the right chemical variation on the theme of layered $CuO_2$ planes would drive transition temperatures upwards. The dream is that one may induce superconductivity at or even above room temperature. Writing in Nature Materials, Hu et al. report [2] an interesting development in this search. Using intense mid-infrared light pump pulses, they resonantly drive large-amplitude (several per cent of the equilibrium lattice constant) oscillations in the out-of-plane apical oxygens (Fig. 1) in the high temperature cuprate superconductor $YBa_2Cu_3O_{6+x}$. By doing so, they create a transient but highly conducting state at temperatures far above the equilibrium superconducting transition temperature. This state develops a feature in the optical reflectivity called the Josephson plasmon resonance, a key indicator of superconductivity at low temperatures. For $YBa_2Cu_3O_{6.45}$, the temperature at which this non-equilibrium effect manifests exceeds room temperature.

The most intriguing interpretation of these observations is that a superconducting, albeit transient, state has been created above room temperature. If true, this would have seminal implications not just for superconductivity, but also for our understanding of non-equilibrium effects in solids. As one may imagine, however, any report of superconductivity above room temperature is not free of controversy. A few considerations and caveats are therefore in order.

Thermal equilibrium is one of the cornerstones of modern statistical mechanics and condensed-matter physics. By considering and averaging over ensembles that do not evolve in time, microscopic physical phenomena can be related to macroscopic physical laws, in a general way that is independent of the specific model used to describe them. Indeed, the idea of thermal equilibrium even allows us to define concepts as basic as that of temperature itself. But there can be reasons for wanting to push physical systems out of equilibrium and study their behaviour. At least three classes of phenomena and motivations for studying a condensed-matter system out of equilibrium come to mind.

Scenario 1 - Experiments are typically performed by 'pumping' the system at one time and 'probing' it at a later stage. After pumping, one hopes to learn something about the relaxation mechanisms and timescales of the equilibrium phase by watching the system relax and return to equilibrium through the decay of its elementary excitations.

Scenario 2 - One may drive the system in such a way as to allow access to material configurations (for example, structure or free charge density) that cannot be accessed in equilibrium. Intense pump pulses may therefore change the free-energy landscape and allow a competing phase to be stabilized in a transient fashion. The hope is that the transient phase reflects equilibrium possibilities in a larger generalized parameter space.

Scenario 3 - One may drive a system to achieve a non-equilibrium phase that cannot exist or is not stabilized without a time - dependent driving field. In this sense, the time-dependent 'pump' should be considered a term in the Hamiltonian. Recent work that claims that a light-illumination-driven 'Floquet' topological insulator can be stabilized is of this variety [3,4].

The observation reported by Hu et al. is at first sight counterintuitive. They blast a delicate superconducting state with light that dumps energy into the system. One would naively expect to suppress superconductivity, not to enhance it. But the cuprates are complicated materials and it is important to consider the three scenarios described above carefully.

The most obvious result of pumping a superconductor with an intense pulse of light is to break Cooper pairs and create a transient conducting state of quasiparticles. There is a large literature on non-resonant pump–probe studies of the cuprates [5] in which quasiparticle decay is examined in the manner of scenario 1. In the case of the present experiment, the authors put forward reasonable arguments suggesting that they are not exciting quasiparticles in this fashion, but this possibility cannot be ruled out completely. However, if it is photoexcited quasiparticles, one would still have to explain the high charge mobilities of the order of 1000 $cm^2$/V.sec observed perpendicular to the $CuO_2$ planes, where transport is expected to be poor and incoherent in the normal state.

In these experiments, the mid-infrared pump pulse does not illuminate the sample indiscriminately. Instead, it drives a phonon which is known to couple strongly to various kinds of spin and charge order, which generally suppress superconducting correlations [6,7,8]. There have been reports of weak superconducting correlations in these compounds to temperatures well above the superconducting transition temperature $T_c$ [9]. A possibility consistent with scenario 2 is that driving the phonon suppresses the competing states and allows the superconductivity to become established. Of course, such an interpretation does not straightforwardly explain the occurrence of superconductivity at room temperature. Still, one cannot exclude the possibility that the transition temperatures of the cuprates would in general be even higher [10] except for the inopportune occurrence of their many competing phases.

An intriguing possibility reflecting scenario 3 is that the effect of driving the phonon actually dynamically stabilizes the superconductivity. Dynamic stabilization is a surprising, but well known phenomenon whereby under rapid periodic changes of an oscillator's parameters it can be stabilized in a configuration in which it might be otherwise unstable. A classic example of such a parametric oscillator is Kapitza's pendulum [11], in which a rigid pendulum can be stabilized with the mass above the pivot point (see Fig. 2). Of course, under ordinary conditions such a configuration is not stable; the weight must hang below the pivot. But if the pivot of the pendulum is made to rapidly oscillate vertically, the behavior of the pendulum changes dramatically. Within a certain range of driving motion, the inverted equilibrium can be stabilized. Although seemingly exotic, parametric oscillation is actually quite a common phenomenon: a child propelling themselves by shifting their center of gravity up and down on a play swing is a prosaic mechanical example. The optical parametric amplifier in which the intense mid-infrared pulses are generated in the present experiment is a related photonic realization of this phenomenon. How dynamic stabilization would work precisely in the present case is unclear, but there is the possibility that it could reduce the thermal fluctuations associated with the c-axis Josephson plasmon and the resulting degradation in the superfluid phase stiffness.

The work of Hu et al. brings up some intriguing possibilities, and future work will hopefully reveal more details about this interesting photoexcited state. It is important to understand for instance the pathways that the energy (~4 mJ per $cm^2$ per pulse) deposited in the system takes.  Time-resolved angle-resolved photoemission could be useful here.  With regards to scenario 2 above, further structural and scattering studies on the photoexcited state would help to establish if fluctuations of a competing state are reduced by photoexcitation. It is important to keep in mind that even if superconductivity has been created at high temperatures, it is only of a transient nature. The most exciting possibility would, of course, be that superconducting correlations (or even fully-fledged superconductivity!) might be stabilized at room temperature in a continuous fashion with a non-pulsed optical system. For now this possibility still seems rather distant, but it's a dream that will keep physicists going for a long while yet.

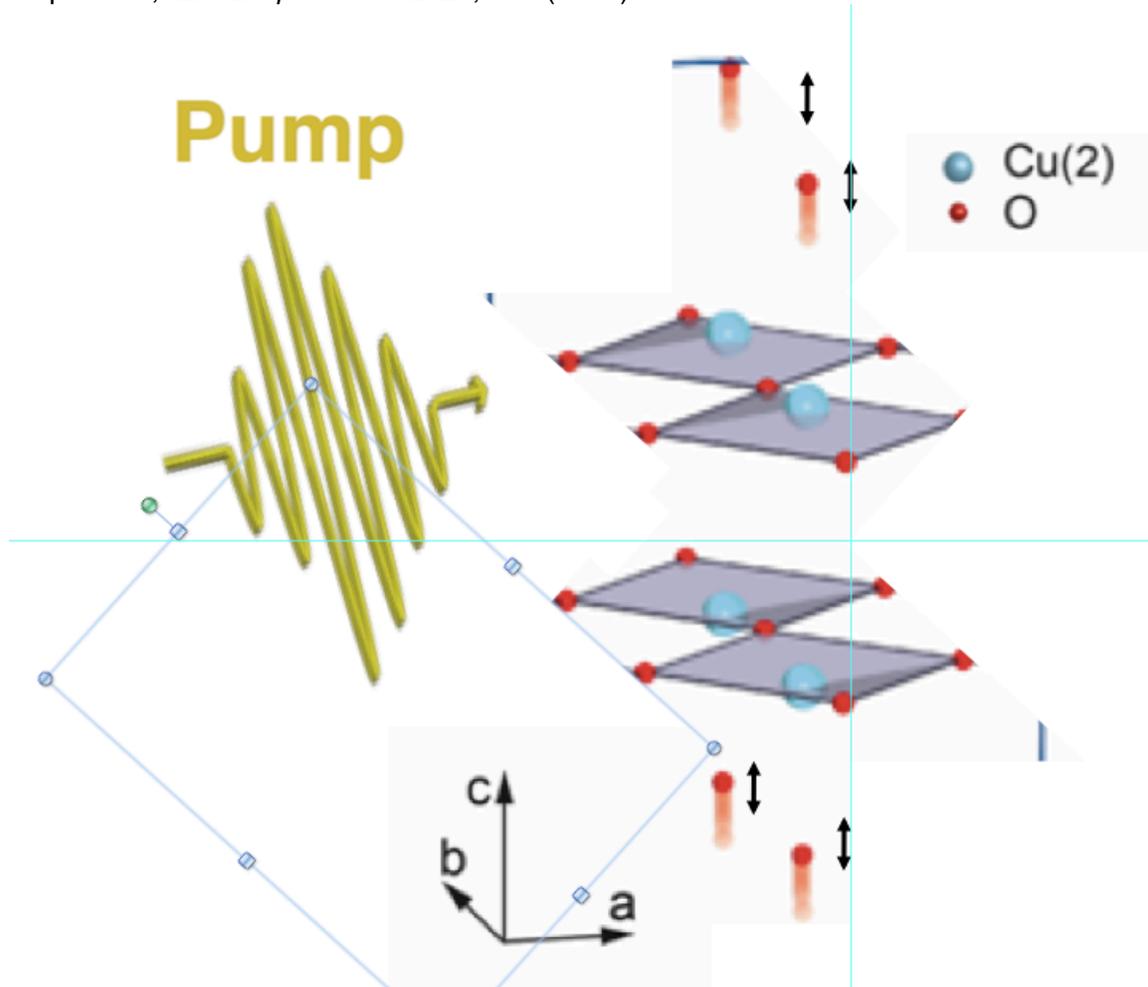

Fig. 1  **Intense light drives oscillations of apical oxygens**.  Hu et al., resonantly drive large-amplitude oscillations of out-of-plane apical oxygens in the high temperature cuprate superconductor $YBa_2Cu_3O_{6+x}$ using intense mid-infrared light pump pulses.  They create a transient, but highly conducting state at temperatures far above the equilibrium superconducting transition temperature.

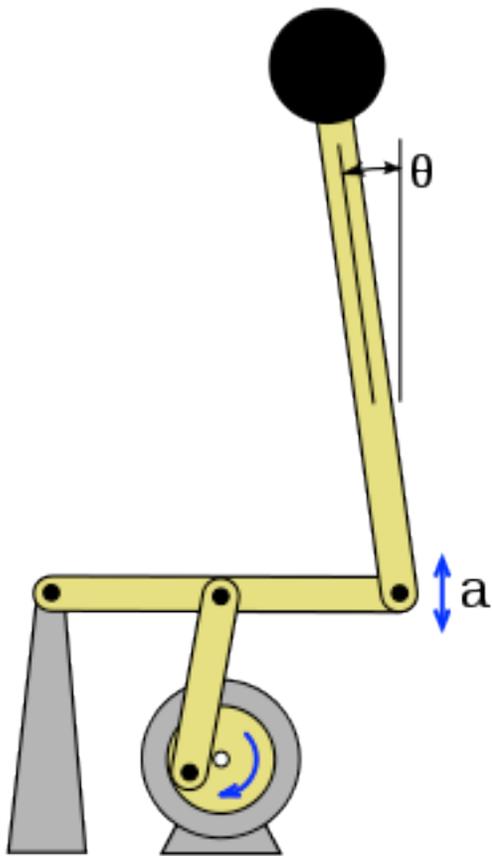

Fig. 2. **A Kapitza pendulum**. This can be constructed with a motor rotating a crank at a high speed. The crank drives a lever arm up and down on which the pendulum is attached to with a pivot. For certain range of driving frequencies, the pendulum is stabilized in the "upside down" configuration.